\documentclass[12pt]{article}
\usepackage[english]{babel}
\usepackage{amsfonts,amsmath,amsthm,graphicx,a4wide}
\usepackage[colorlinks=true]{hyperref}
\usepackage[]{authblk}

\newcommand{\dotex}{\frac{d}{dt}}
\newcommand{\tr}[1]{\text{Tr}\left(#1\right)}

\newcommand{\RR}{{\mathbb R}}

\newcommand{\bra}[1]{\langle #1 |}
\newcommand{\ket}[1]{| #1 \rangle}

\newcommand{\ba}{\text{\bf{a}}}
\newcommand{\bn}{\text{\bf{n}}}

\newcommand{\Id}{{\mathbb I}}
\newcommand{\Dm}{{\mathcal D}_{m}}
\newcommand{\Pm}{\Pi_{m}^{\rho}}
\newcommand{\dt}{\delta\hspace{-0.08em}t}

\newcommand{\dO}{\delta\hspace{-0.1em}O}
\newcommand{\dR}{\delta\hspace{-0.1em}R}

\newcommand{\dH}{\eta}
\newcommand{\ds}{\varsigma}
\newcommand{\dr}{\xi}
\newcommand{\kd}{{k\!+\!\text{\tiny$\tfrac{1}{2}$}}}
\newcommand{\kt}{{k\!+\!\text{\tiny$\tfrac{1}{3}$}}}
\newcommand{\ktt}{{k\!+\!\text{\tiny$\tfrac{2}{3}$}}}

\title{Low rank approximation for the numerical simulation of high dimensional Lindblad and Riccati equations}
\author{ C. Le Bris $^1$ \& P. Rouchon $^2$ \medskip \\
{\footnotesize $^1$ \'Ecole Nationale des Ponts et Chauss\'ees,}\\
{\footnotesize 6 et 8 avenue Blaise Pascal, 77455 Marne-La-Vall\'ee Cedex 2, France}\\
{\footnotesize  and INRIA Rocquencourt, MICMAC project-team,Domaine de Voluceau, B.P. 105,}\\
{\footnotesize  78153 Le Chesnay Cedex, FRANCE.}\\
{\footnotesize\tt lebris@cermics.enpc.fr}\\
{\footnotesize $^2$ Mines-ParisTech, Centre Automatique et Syst\`{e}mes, 60, bd Saint-Michel, 75006 Paris, France. }\\
{\footnotesize \tt pierre.rouchon@mines-paristech.fr}
}

\begin{document}
\maketitle
\begin{abstract}
A systematic  numerical  approach to
approximate  high dimensional Lindblad  equations is described. It is based on  a deterministic
rank $m$  approximation of the density operator, the rank $m$ being the only parameter to adjust. From a known initial
value, this rank $m$ approximation   gives  at each time-step   an estimate of the largest $m$ eigen-values with their   eigen-vectors of the density operator. A numerical scheme is proposed. Its numerical efficiency in the case of a   rank $m=12$ approximation  is demonstrated for oscillation revivals of $50$ atoms interacting resonantly  with a slightly
damped  coherent quantized field of $200$ photons. The approach may be employed for other similar equations. We in particularly show how to adapt such low-rank  approximation for Riccati differential equations appearing in Kalman filtering.
\end{abstract}
\section{Introduction}

Numerical simulations of high dimensional Lindblad equations are routinely
performed using  ensemble averages of quantum Monte-Carlo
trajectories~\cite{dalibard-et-al:PRL92,molmer-et-al:JOSA93,haroche-raimond:book06}. We propose here another
approach that can be also adapted to high dimensional  matrix Riccati equations. This approach consists in approximating the evolution of the $n\times n$ density
matrix $\rho$  solution to the differential Lindblad equation using a
reduced dynamics on the set of  density matrices of some fixed rank $m <
n$. This reduced dynamics  is obtained by  taking  the orthogonal projection  of $\dotex \rho$ onto the tangent space to this set of matrices of rank $m$.

Such an  approximation strategy,  based on  orthogonal projections onto
low dimensional manifolds,   has  already been proposed
in~\cite{Handel-mabuchi:JOB2005,Mabuchi:PRA2008} in the context of
quantum filtering. The goal then was to construct  a reduced order
quantum filter for a spin-spring system. The  submanifold on which the
dynamics was projected was the (real) 4-dimensional manifold constructed with the  tensor
products of  arbitrary two-level states and   pure coherent
states. In~\cite{bonnabel-sepulchre:2012}, a geometric  approach to
study the stability of low-rank solutions  for   matrix  Riccati  differential equations has been likewise developed.
We combine here the ideas of~\cite{Handel-mabuchi:JOB2005}
and~\cite{bonnabel-sepulchre:2012}. We focus on the  Lindblad
case. The Riccati case  is addressed in appendix~\ref{ap:riccati} without numerical simulations. Nevertheless, in both cases, the proposed  approximation  depends only on one adjustable parameter, the low-order rank~$m$.

When the size $n\times n$ of the density matrix
exceeds the computing possibilities available --which is often the
case in practice even for rather simple physically relevant systems--, such an approximation
can be very useful to compute  the (approximate) time evolution of $\rho$ from a known initial
value with rank smaller than or equal to  $m$.  This approximation strategy  is
tested here  on  oscillation revivals of $N_a$ atoms in a slightly damped  mesoscopic
field  with $\bar n$ photons in
average~\cite{Meunier-et-al:PRA2006}. In
a first stage, we consider the original half-spin/spring
case~\cite{Gea-Banacloche:PRA1991} with $N_a=1$ and $\bar n= 15$. For a
photon  lifetime in the order of one hundred  vacuum Rabi periods, we compare  the numerical solution  $\rho(t)$ to the  Lindblad
equation (computed for reference) with  different numerical low-rank solutions from $m=2$ to
$6$. Our results show a very satisfactory agreement. We next take the much larger values $N_a=50$ and $\bar n =200$. The
density matrix is then too large for an explicit comparison with the direct numerical integration of
the Lindblad differential equation  to be possible using a standard
computer ($n\sim 15000$).  For a photon lifetime in the order of ten thousands vacuum Rabi periods, we have checked that our numerical solution
using a reduced model of  rank $m=12$ is consistent with  the  analytic damping model proposed in~\cite{Meunier-et-al:PRA2006}. We also observe  numerically that the larger the
rank $m$, the more accurate the approximation.

Our article is articulated as follows. We derive in
Section~\ref{sec:proj} the low rank dynamics approximating the
Lindblad equation. We describe the numerical
scheme we use in Section~\ref{sec:num}. In both sections, our general
purpose strategy
is specifically adjusted to account for the fact that, usually, the Hamiltonian part dominates the
decoherence part. Section~\ref{sec:revival} presents our numerical
experiments on oscillation revivals. We draw some conclusions in Section~\ref{sec:conclusion}. In appendix~\ref{ap:riccati}, a similar  low-rank approximated dynamics   is derived  for  the matrix Riccati equation appearing in Kalman filtering. Such low-rank approximates of the covariance matrix could be eventually useful in data assimilation for large scale systems as those  governed by partial differential equations.

\paragraph{Acknowledgements} The authors thank  Silv\`{e}re Bonnabel,
Michel Brune, Michel Devoret, Jean-Michel Raimond for useful references
and stimulating interactions. The authors were partially supported by the ANR, Projet Blanc EMAQS ANR-2011-BS01-017-01.

\section{The Low rank differential  equations} \label{sec:proj}
Consider a Lindblad equation with, for simplicity (see last paragraph of this section), a single decoherence operator $L$,
\begin{equation}\label{eq:lindblad}
    \dotex \rho = -i[H,\rho]  - \tfrac{1}{2} (L^\dag L \rho + \rho L^\dag L) + L \rho L^\dag,
\end{equation}
where $\rho$ is $n\times n$ non-negative Hermitian matrix with
$\tr{\rho}=1$, $H$ is a $n\times n$ Hermitian matrix and $L$ is a $n\times
n$ matrix. Our purpose is to approximate, for $n$ large,  the above
dynamics on the set of non-negative Hermitian matrices of rank $m$, $m$
being an integer presumably much
smaller than $n$, prescribed beforehand. We now formalize this.

\medskip

A density $\rho$ of rank $m< n$ can be decomposed as
\begin{equation}\label{eq:sigma}
\rho = U \sigma U^\dag
\end{equation}
where $\sigma$ is a $m\times m$ strictly positive Hermitian matrix,  $U$
a $n\times m$ matrix with $U^\dag U =\Id_m$, and $\Id_m$ of course
denotes the $m\times m$ identity matrix.
The set of  non-negative Hermitian  matrices of rank $m$ and trace one
can be seen as a sub-manifold, denoted by $\Dm$,  of the Euclidean
space of $n\times n$ Hermitian matrices equipped with the Frobenius
scalar product. For $\rho\in\Dm$, $\dotex \rho$ given
by~\eqref{eq:lindblad} does not belong in general to the tangent space
to $\Dm$  at~$\rho$. The rank is therefore not necessarily preserved by
the evolution governed by~\eqref{eq:lindblad}. To correct for this, the
most natural option is to
consider, for any $\rho\in\Dm$,  the orthogonal projection of
$\dotex\rho$ given by~\eqref{eq:lindblad} onto the tangent space to
$\Dm$ at $\rho$. Denoting by  $\Pm$ the projection operator, we
therefore consider the differential equation
 \begin{equation}\label{eq:LowRank}
    \dotex \rho = \Pm\left(-i[H,\rho]  - \tfrac{1}{2} (L^\dag L \rho + \rho L^\dag L) + L \rho L^\dag\right)
\end{equation}
set on $\Dm$, which can be seen as a
rank $m$ approximation of~\eqref{eq:lindblad}.

Making  the approach practical requires to now give an explicit formulation
of  such a rank $m$ approximation.  A  lifting procedure,  inspired
from~\cite{bonnabel-sepulchre:2012} and described in some more details
below, consists in introducing two coupled  differential equations for $U$ and $\sigma$ corresponding to
the generic decomposition~\eqref{eq:sigma} for $\rho\in\Dm$:
\begin{align}
\dotex U &= - i A U + (\Id_n-UU^\dag) \left(-i(H-A) - \tfrac{1}{2}L^\dag L +LU\sigma U^\dag L^\dag U\sigma^{-1} U^\dag\right) U,
    \label{eq:U}
    \\
  \dotex \sigma &= - i[U^\dag (H-A) U,\sigma] - \tfrac{1}{2} (U^\dag L^\dag L U \sigma + \sigma U^\dag L^\dag L U)
  + U^\dag L U\sigma  U^\dag L^\dag U \notag
  \\&\qquad + \tfrac{1}{m}\tr{(L^\dag (\Id_n-UU^\dag)L ~U\sigma U^\dag} \Id_m.
    \label{eq:sig}
\end{align}
In \eqref{eq:U}-\eqref{eq:sig},  $A$ denotes an  arbitrary Hermitian
operator that may depend on time. Then  standard manipulations, which we
omit here for brevity, show that
\begin{itemize}
  \item $U^\dag U$ remains equal to $\Id_m$,
  \item $\sigma$ remains Hermitian, positive and of trace one,
\end{itemize}
and that
the evolution of  $\rho= U\sigma U^\dag $ then solves~\eqref{eq:LowRank}, which in this particular instance reads
\begin{multline}\label{eq:LowRankbis}
    \dotex \rho =  -i[H,\rho]  - \tfrac{1}{2} (L^\dag L \rho + \rho L^\dag L) + L \rho L^\dag - (\Id_n-P_\rho)L\rho L^\dag (\Id_n-P_\rho) \\+\tfrac{\tr{L\rho L^\dag(\Id_n-P_\rho)}}{m} P_\rho,
\end{multline}
where $P_\rho=UU^\dag $ only depends on $\rho$ since it corresponds to  the orthogonal projection on the image of $\rho$.

Notice that~\eqref{eq:LowRankbis} does not depend
on the arbitrary matrix $A$: its entries  can be seen as gauge degrees
of freedom. The specific choice $A=H$ yields
\begin{align}
\dotex U &= - i H U + (\Id_n-UU^\dag) \left(- \tfrac{1}{2}L^\dag L +LU\sigma U^\dag L^\dag U\sigma^{-1} U^\dag\right) U,
    \label{eq:UH}
    \\
  \dotex \sigma &=  - \tfrac{1}{2} (U^\dag L^\dag L U \sigma + \sigma U^\dag L^\dag L U)
  + U^\dag L U\sigma  U^\dag L^\dag U \notag
  \\&\qquad + \tfrac{1}{m}\tr{(L^\dag (\Id_n-UU^\dag)L ~U\sigma U^\dag} \Id_m
  ,
    \label{eq:sigH}
\end{align}
where we note that  $H$ only appears in the dynamics for $U$ and not in the dynamics of~$\sigma$, a
choice that is particularly appropriate when $H$ dominates $L$. In that
case indeed,
\eqref{eq:sigH} may be understood as a slow evolution as compared to the
 dynamics~\eqref{eq:UH}. The efficiency of our numerical procedure,
 described in the next section, will significantly benefit from this
 particular decomposition.

\medskip

We now explain how we obtain~\eqref{eq:U} and~\eqref{eq:sig}. A $n\times n$ Hermitian matrix  $\dr$  in the   tangent space at  $\rho=U \sigma U^\dag $ to $\Dm$  admits the parameterization
\begin{equation}\label{eq:drho}
    \dr = i[\dH,\rho] + U \ds U^\dag = U \big( i[U^\dag \dH U, \sigma] + \ds \big) U^\dag
\end{equation}
where $\dH$ is any  $n\times n$ Hermitian matrix and $\ds$   any
$m\times m$ Hermitian matrix with zero trace.  This results from the
definition of the tangent map of the submersion
$(U,\sigma) \mapsto U\sigma U^\dag$ with the infinitesimal variations  $\delta U = \imath \dH U$ and
$\delta\sigma = \ds$. The parameterization~\eqref{eq:drho} is onto, but
not one-to-one: note that different $\dH$ and $\ds$ may indeed yield
the same $\dr$. The projection
$\Pm(\dotex\rho)$ corresponds to the tangent vector $\dr$ associated  to  $\dH$ and $\ds$ minimizing
$$
\tr{\Big(S\rho+\rho S^\dag + L\rho L^\dag - i[\dH,\rho] - U \ds U^\dag\Big)^2 },
$$
where $S=-iH- L^\dag L /2$.
The first order stationary  conditions versus $\ds$ and $\dH$ read
\begin{align*}
    U^\dag \big( S \rho + \rho S^\dag  + L \rho L^\dag - i[\dH,\rho]- U\ds U^\dag \big) U &= \lambda \Id_m\quad,
    \\
    \big[S \rho + \rho S^\dag + L \rho L^\dag  - i[\dH,\rho]- U\ds U^\dag,\rho \big]=0,
\end{align*}
where the real scalar $\lambda$ is implicitly given by the constraint $\tr{\ds}=0$.
The first equation yields
\begin{equation}\label{eq:ds}
\ds= U^\dag \big( S \rho + \rho S^\dag + L\rho L^\dag  - i[\dH,\rho] \big) U - \tfrac{1}{m}\tr{U^\dag(S\rho+\rho S^\dag+L \rho L^\dag )U} \Id_m.
\end{equation}
We insert this value of $\ds$ into the second equation and obtain,  after
some easy manipulations using $U^\dag U = \Id_m$,
$$
(\Id_n-P) \big( S \rho +L\rho L^\dag - i  \dH \rho \big)\rho
=
\rho \big(\rho S^\dag   + L\rho L^\dag  + i \rho \dH \big) (\Id_n-P)
$$
where $P=U U^\dag$ is the orthogonal projector on the image of $U$.
This matrix equation admits the following general solution
\begin{multline}\label{eq:dH}
    \dH = - i (\Id_n-P) \big( S P + L \rho L^\dag U \sigma^{-1} U^\dag \big)
    + i \big( P S^\dag  + U \sigma^{-1} U^\dag L \rho L^\dag \big)(\Id_n-P)
    \\
    - P A P - (\Id_n-P)A(\Id_n-P)
\end{multline}
where $A$ is an  arbitrary Hermitian operator.
Then, using $U^\dag P = U^\dag$ and $PU=U$, $\ds$ given by~\eqref{eq:ds} reads
\begin{multline}\label{eq:ds_bis}
    \ds =  - i[U^\dag H U,\sigma] - \tfrac{1}{2} (U^\dag L^\dag L U \sigma + \sigma U^\dag L^\dag L U)
  + U^\dag L U\sigma  U^\dag L^\dag U
  \\ + \tfrac{1}{m}\tr{(L^\dag (\Id_n-UU^\dag)L ~U\sigma U^\dag} \Id_m + i [U^\dag A U,\sigma]
\end{multline}
The derivatives $\dotex U = \imath \dH U$ and
$\dotex\sigma = \ds$ respectively give~\eqref{eq:U} and~\eqref{eq:sig}.

\medskip

 We finally note that, for
 clarity, we  have deliberately restricted our exposition to the case
 of a single decoherence  term in~\eqref{eq:lindblad}.  The
 generalization to an arbitrary number of decoherence terms
 $$
 \dotex \rho = -i[H,\rho] + \sum_{\nu} L_\nu \rho L_\nu^\dag - \tfrac{1}{2} (L_\nu^\dag L_\nu \rho + \rho L_\nu^\dag L_\nu)
 ,
 $$
 is straightforward: \eqref{eq:UH} and \eqref{eq:sigH} become
 \begin{align*}
\dotex U &= - i H U + (\Id_n-UU^\dag) \left(\sum_\nu - \tfrac{1}{2}L_\nu^\dag L_\nu +L_\nu U\sigma U^\dag L_\nu^\dag U\sigma^{-1} U^\dag\right) U
    \\
  \dotex \sigma &=  \sum_\nu \tfrac{-1}{2} (U^\dag L_\nu^\dag L_\nu U \sigma + \sigma U^\dag L_\nu^\dag L_\nu U)
  + U^\dag L_\nu U\sigma  U^\dag L_\nu^\dag U
  \\&\qquad + \tfrac{1}{m}\tr{\sum_\nu(L_\nu^\dag (\Id_n-UU^\dag)L_\nu ~U\sigma U^\dag} \Id_m
  .
\end{align*}

\section{Numerical integrator}
\label{sec:num}

 For  a positive integer $k$ and  the timestep $\dt$, we denote by $U_k$
 and $\sigma_k$ the numerical approximations of $U(k\dt)$ and
 $\sigma(k\dt)$ solutions to~\eqref{eq:UH} and~\eqref{eq:sigH} respectively.  The
 evolution from time $k\dt$ to time $(k+1)\dt$ is split into the following  three steps:
\begin{itemize}

\item The first step consists in an approximation of the free
  Hamiltonian evolution  $U_{\kt}=e^{-(i\dt/2) H} U_k$. For
  the simulations of Section~\ref{sec:revival}, we choose  a (formal)
  third-order expansion:
\begin{equation}\label{eq:Ukt}
    U_{\kt} = \left(\Id_n - \tfrac{i \dt}{2} H - \tfrac{\dt^2}{8} H^2+ i \tfrac{\dt^3}{48} H^3 \right) U_k.
\end{equation}
We however note  that, of course, other choices
  are possible. In particular, time integrators more adapted to the time
  discretization of the Schr\"odinger may be employed
  (see~\cite{lubich-book} and references therein). We will not proceed in this direction in the
  present work and,  should
  need be, hope to return to this issue in future works.
\item The second step consists in updating both $U$ and $\sigma$ now
  accounting for $L$; we set
\begin{equation}\label{eq:Uktt}
    U_{\ktt} =  U_{\kt} + \dt (\Id_n-U_{\kt}U_{\kt}^\dag) \left(- \tfrac{1}{2}L^\dag L U_{\kt}  +LU_{\kt}\sigma_k U_{\kt}^\dag L^\dag U_{\kt}\sigma_k^{-1} \right)
\end{equation}
and, in two stages,
\begin{eqnarray}
    \sigma_{\kd}&= &\sigma_k + \dt~  U_{\kt}^\dag L U_{\kt}\sigma_k
    U_{\kt}^\dag L^\dag U_{\kt} \nonumber\\
&&+ \dt~\frac{\tr{( U_{\kt}^\dag L^\dag L U_{\kt} - U_{\kt}^\dag L^\dag U_{\kt} U_{\kt}^\dag  L U_{\kt}) \sigma_k} ~\Id_m}{m} ,\label{eq:sigkd}
\end{eqnarray}
followed by
\begin{equation}\label{eq:sigkp1}
    \sigma_{k+1}= \frac{(\Id_m- \tfrac{\dt}{2} U_{\kt}^\dag L^\dag L U_{\kt})  \sigma_\kd (\Id_m- \tfrac{\dt}{2} U_{\kt}^\dag L^\dag L U_{\kt})}
    {\tr{(\Id_m- \tfrac{\dt}{2} U_{\kt}^\dag L^\dag L U_{\kt})  \sigma_\kd (\Id_m- \tfrac{\dt}{2} U_{\kt}^\dag L^\dag L U_{\kt})}}
    .
\end{equation}
We note that this two-stage update corresponds to a slight modification
of  the explicit Euler scheme so that it preserves both positiveness and
the trace. In particular, the motivation for using~\eqref{eq:sigkp1} can
be understood from the following considerations (which can be
adapted to many similar contexts). Momentarily omitting its
rightmost term for simplicity, we observe that, with obvious notation, \eqref{eq:sigH} is of the form
$\displaystyle \dotex \sigma
=W\,\sigma+\sigma\,W^\dag+Z\,\sigma\,Z^\dag$. Introducing the auxiliary
variables~$\tilde\sigma=B\,\sigma\,B^\dag$
and~$\tilde Z=B\,Z\,B^{-1}$ with $B$ solution to
$\displaystyle\dotex B=-BW$, we see that $\displaystyle \dotex \tilde\sigma=\tilde Z\,\tilde \sigma\,\tilde Z^\dag$, an equation that can be simply
integrated using the first order forward Euler scheme
$\tilde\sigma_{k+1}=\tilde\sigma_{k+1/2}+\delta t/2\,\,\tilde Z_{k+1/2}\,\tilde
\sigma_{k+1/2}\,\tilde Z_{k+1/2}^\dag$. Expressed in the original unknown~$\sigma$,
this scheme gives the numerator of~\eqref{eq:sigkp1} up to $\dt^2$ terms and automatically preserves
positiveness. The preservation of the trace is then ensured by the
normalization in~\eqref{eq:sigkp1}.

\item The third step is similar to the first step, the  third-order
  approximation of  $U_{k+1}=e^{-(i\dt/2) H} U_{\ktt}$, this time followed by an
  orthonormalization (a procedure formally  denoted by  $\Upsilon$) to ensure $U_{k+1}^\dag U_{k+1} = \Id_m$:
\begin{equation}\label{eq:Ukp1}
    U_{k+1} = \Upsilon \Big( \left(\Id_n - \tfrac{i \dt}{2} H - \tfrac{\dt^2}{8} H^2+ i \tfrac{\dt^3}{48} H^3  \right) U_{\ktt} \Big)
    .
\end{equation}

\end{itemize}
\medskip

Three comments, different in nature, are in order.

\paragraph{Computational cost} The above numerical scheme  essentially
uses  right  multiplications of  $H$, $L$, $L^\dag$   by $n\times m$ matrices,
as, for example, the products  $HU$, $H^2U=H(HU)$,  $LU$, $L^\dag (L U)$. To evaluate such
products does not require string $n\times n$ matrices since usually $H$
and $L$ are defined as tensor products of operators of small
dimensions. When $n$ is very large and $m$ is small, this point is
crucial for an efficient numerical implementation of the approach. In particular evaluations of  products like $HU$ or $LU$ can be easily parallelized.

\paragraph{Formal error estimator}
To obtain an empirical  estimate
of the error
committed when using the low rank approximation, we may
follow~\cite{Handel-mabuchi:JOB2005} and compute the Frobenius norm of $\dot\rho=\dotex \rho $ and  $ \dot\rho_\perp = \dot\rho-\Pm(\dot\rho)$ at each time step.
 If the Frobenius norm of
$\dot\rho_\perp$  is much smaller than that of $\dot\rho$, then the
approximation is considered valid.
From~\eqref{eq:LowRankbis}, we have  the following general formulae for
any $\rho\in\Dm$,  any Hermitian operator $H$  and any  not necessarily
Hermitian operator $L$:
\begin{align*}
    \Pm \Big([H,\rho]\Big) &= [H,\rho] \notag
    \\
    \Pm \Big(- \tfrac{1}{2} (L^\dag L \rho + \rho L^\dag L) + L \rho L^\dag\Big) &=
   - \tfrac{1}{2} (L^\dag L \rho + \rho L^\dag L) + L \rho L^\dag\notag \\& \qquad - (\Id_n-P_\rho)L\rho L^\dag (\Id_n-P_\rho) +\tfrac{\tr{L\rho L^\dag(\Id_n-P_\rho)}}{m} P_\rho
    .
%     \label{eq:OrthProj}
\end{align*}
Thus we have,
\begin{equation}\label{eq:rhoperp}
 \dot\rho_\perp = (\Id_n-P_\rho)L\rho L^\dag (\Id_n-P_\rho) - \tfrac{\tr{L\rho L^\dag(\Id_n-P_\rho)}}{m} P_\rho
\end{equation}
where $P_\rho=UU^\dag$ since $\rho = U\sigma U^\dag$. Classical
computations using standard properties of the trace show that, at each
time step, $\tr{\dot \rho^2}$ and
$\tr{\dot\rho_\perp^2}$ may be numerically evaluated with a complexity similar to the
complexity of the  numerical scheme. There is no need to explicitly compute  $\dot\rho $ and $\dot\rho_\perp$ as  $n\times n$ matrices before taking their Frobenius norm.

\paragraph{Choice of the initial condition} Initial conditions for
$\sigma$ and $U$ need to be deduced  from a given initial
condition $\rho_0$. When the rank
of $\rho_0$ is larger than  or equal to $m$, we form  $\sigma_0$ as the diagonal matrix consisting of the
largest $m$  eigenvalues of $\rho_0$ with sum normalized to one and we
form $U_0$   using the associated
eigenvectors. When the rank
of $\rho_0$ is strictly less than  $m$ then we proceed as follows.

Assume in a first step  that $\rho_0$ is a pure state $\ket{\psi_0}\bra{\psi_0}$. It is then natural to take for  $\sigma_0$  a  diagonal matrix where the first diagonal element is $1-(m-1)\epsilon$ and the over ones are equal to  $\epsilon$. Here  $\epsilon$ is a positive number much smaller than $1$ (typically $10^{-5}$). Then  $U_0$ is constructed, up to an orthonormalization preserving the first column,  with $\ket{\psi_0}$ as the first  column, $H\ket{\psi_0}$ as the second column, \ldots, $H^{m-1}\ket{\psi_0}$ as the last column.

 When the rank of $\rho_0$ is  strictly larger than $1$ (but still strictly
 less than $m$), one can easily imagine the following mixed procedure: put the non-zero eigenvalues in the first  diagonal elements of $\sigma_0$, their associated eigenvectors in the first columns of  $U_0$,  set  the remaining diagonal elements of $\sigma_0$ to $\epsilon$  and  complete  the remaining columns of $U_0$ by iterates of $H$  on these eigenvectors.

\section{Numerical tests for oscillation revivals} \label{sec:revival}

The collective  behavior of $N_a$ atoms resonantly interacting with a quantized  single-mode field  trapped in an almost perfect cavity  is modeled by the following Lindblad master  equation (see, e.g., \cite{Meunier-et-al:PRA2006})
\begin{equation}\label{eq:Jaynes-Cummings-Natoms}
\dotex \rho = \frac{\Omega_0}{2} [ \ba^\dag J^- - \ba J^+, \rho]  - \kappa ( \bn \rho/2+\rho \bn/2 - \ba \rho \ba^\dag )
\end{equation}
where $\Omega_0 >0$ is the coupling strength (vacuum Rabi pulsation), $\ba$ the field annihilation operator, $\bn=\ba^\dag \ba$ the field photon-number operator and $\kappa>0$ the inverse of the single photon life-time. The   $N_a$ atoms  are described here with the spin $J=N_a/2$ representation. It involves  the $N_a+1$  Dicke states $\ket{J,-N_a/2}$, $\ket{J,(-N_a+1)/2}$, \ldots, $\ket{J, N_a/2}$.
These Dicke states are  labelled  by the index $\mu=0$ to $\mu=N_a$ and are denoted by $\ket{\mu}$: for $\mu=0$ all atoms are in the same ground state; for $\mu=N_a+1$ all atoms are in the same  excited state. With this notation, the atomic lowering operator, $J^{-}=(J^{+})^\dag$, is given by  $J^{-}\ket{\mu}= \sqrt{\mu(N_a-\mu)} \ket{\mu-1}$, $\mu=0,\ldots, N_a$.

 For $N_a=1$ atom initially in the excited state, a  field  initially in
 a coherent state with $\bar n=15$ photons (truncation to $30$ photons),
 and a damping factor $\kappa=\Omega_0/500$,  we have compared,  until
 the first revival appearing around   $t =
 \frac{2\pi}{\Omega_0/2\sqrt{\bar n}}$,   low-rank   trajectories,
 solutions to~(\ref{eq:UH},\ref{eq:sigH}),    with the original
 full-rank trajectory solution to~\eqref{eq:lindblad} where  $H=i
 \frac{\Omega_0}{2} (\ba^\dag J^- - \ba J^+)$ and $L=\sqrt{\kappa}
 \ba$. In this  simple case performing the full-rank simulation for
 the sake of comparison is
 indeed feasible. Initializations of $U$ and $\sigma$ are performed according to the procedure  explained at the end of Section~\ref{sec:num}.

\begin{figure}
\centerline{\includegraphics[width=1.2\textwidth]{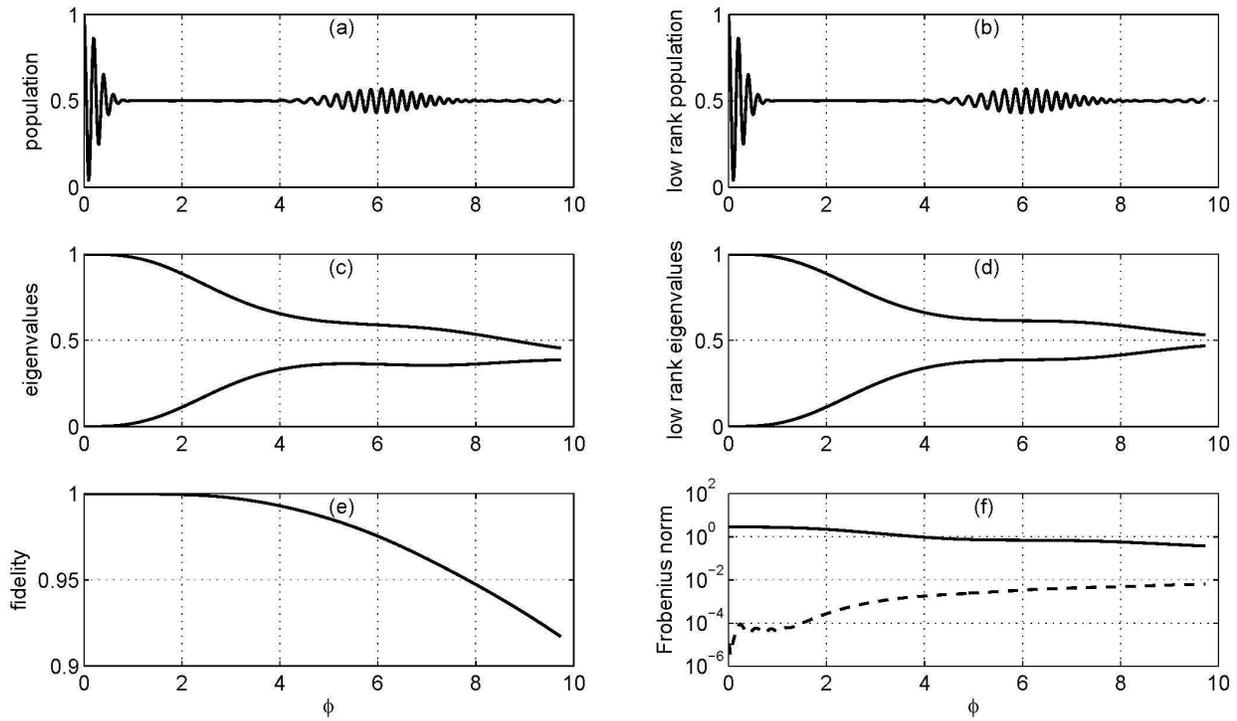}}
  \caption{Oscillation revival for $N_a=1$ and a coherent initial field with $\bar n=15$ photons; numerical solutions of  original  Lindblad master equation~\eqref{eq:lindblad} and  of  the   rank 2 approximation~(\ref{eq:UH},\ref{eq:sigH})  with the adimensionalized time $\phi=\Omega_0 t/2\sqrt{\bar n}$; (a) and (b) correspond to the excited atomic populations; (c) and (d) show the first $2$  eigenvalues; (e) is the fidelity between the Lindblad solution and the rank 2 solution; in (f) the solid  and dashed  curves correspond respectively to the Frobenius norms of $\dot\rho$ and $\dot\rho_\perp$ as explained in~\eqref{eq:rhoperp}.  }\label{fig:HalfSpinRank2}
\end{figure}

\begin{figure}
\centerline{\includegraphics[width=1.2\textwidth]{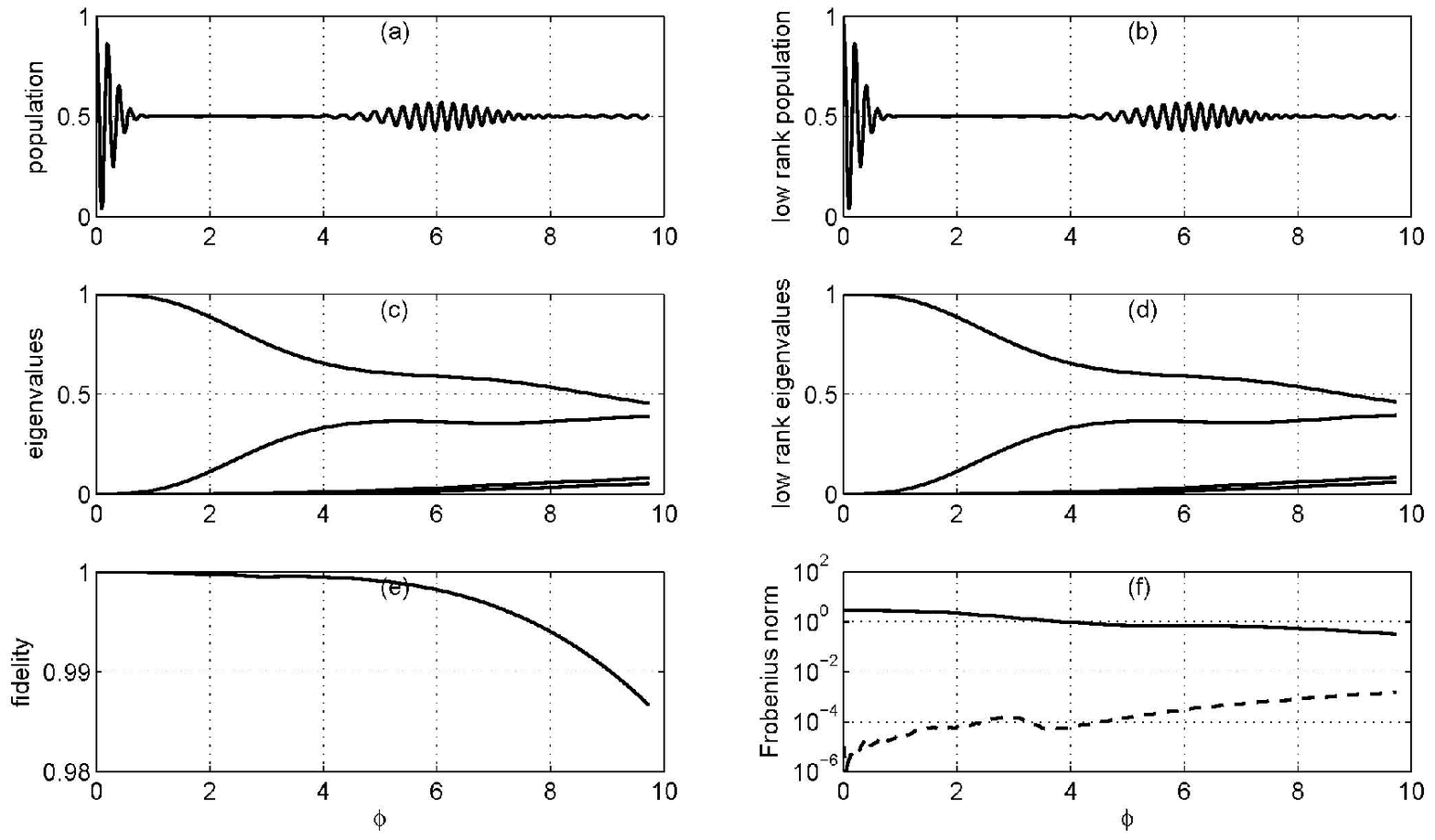}}
  \caption{Oscillation revival for $N_a=1$ and a coherent initial field with $\bar n=15$ photons; numerical solutions of  original  Lindblad master equation~\eqref{eq:lindblad} and  of  the   rank 4 approximation~(\ref{eq:UH},\ref{eq:sigH})  with the adimensionalized time $\phi=\Omega_0 t/2\sqrt{\bar n}$; (a) and (b) correspond to the excited atomic populations; (c) and (d) show the first $4$  eigenvalues; (e) is the fidelity between the Lindblad solution and the rank 4 solution; in (f) the solid  and dashed  curves correspond respectively to the Frobenius norms of $\dot\rho$ and $\dot\rho_\perp$ as explained in~\eqref{eq:rhoperp}.  }\label{fig:HalfSpinRank4}
\end{figure}

\begin{figure}
\centerline{\includegraphics[width=1.2\textwidth]{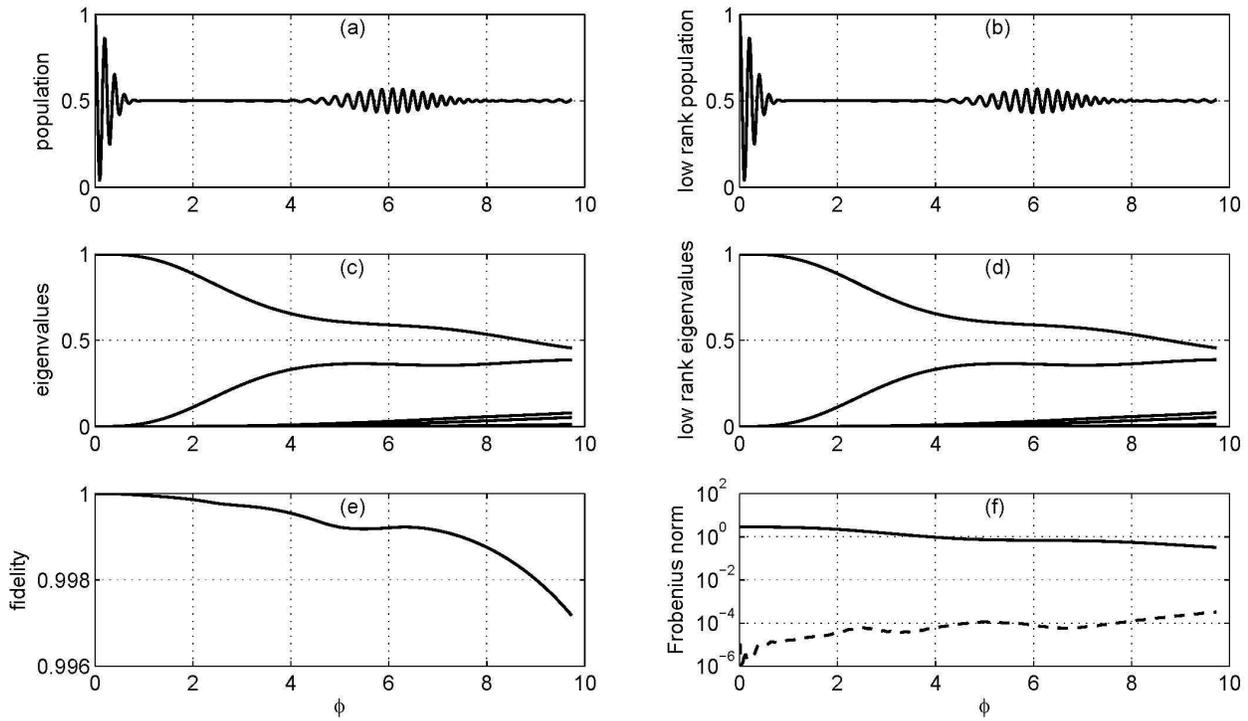}}
  \caption{Oscillation revival for $N_a=1$ and a coherent initial field with $\bar n=15$ photons; numerical solutions of  original  Lindblad master equation~\eqref{eq:lindblad} and  of  the   rank 6 approximation~(\ref{eq:UH},\ref{eq:sigH})  with the adimensionalized time $\phi=\Omega_0 t/2\sqrt{\bar n}$; (a) and (b) correspond to the excited atomic populations; (c) and (d) show the first $6$  eigenvalues; (e) is the fidelity between the Lindblad solution and the rank 6 solution; in (f) the solid  and dashed  curves correspond respectively to the Frobenius norms of $\dot\rho$ and $\dot\rho_\perp$ as explained in~\eqref{eq:rhoperp}.  }\label{fig:HalfSpinRank6}
\end{figure}

Figures~\ref{fig:HalfSpinRank2}, \ref{fig:HalfSpinRank4}
and~\ref{fig:HalfSpinRank6} show that ranks $m\geq 4$  ensure a fidelity
higher than $98\%$ over the interval of simulation, and that the fidelity
increases with the rank.  This is corroborated by the fact that, for
$m=4$ and $m=6$, the Frobenius norm of $\dot\rho_\perp$ is always  less
than $1\%$ of the Frobenius norm of $\dot\rho$. We recall that the
\emph{fidelity} between two density operators  $\rho_a$ and $\rho_b$  is defined by $ \tr{\sqrt{\sqrt{\rho_a}\rho_b\sqrt{\rho_a}}}. $


\clearpage

\begin{figure}
\centerline{\includegraphics[width=1.\textwidth]{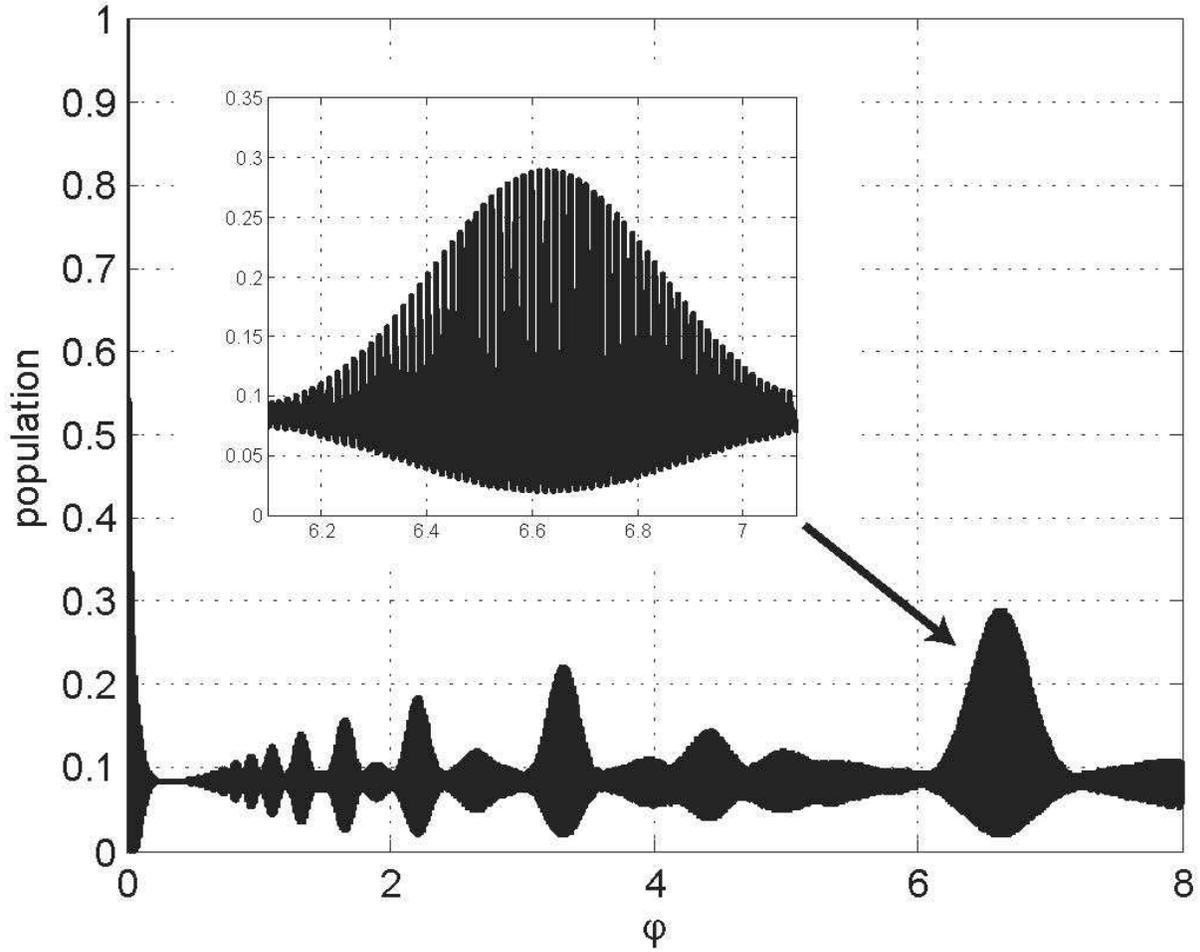}} % simu time 1h15
  \caption{Oscillation revival for $N_a=50$ and a coherent initial field with $\bar n=200$ photons; evolution with the adimensionalized time $\phi=\Omega_0 t/2\sqrt{\bar n}$ of the excited atomic  population (all atoms in the same  excited state); numerical solution to the Schr\"{o}dinger equation~\eqref{eq:Jaynes-Cummings-Natoms} with  $\kappa=0$.  }
  \label{fig:Nspin}
\end{figure}

\begin{figure}
\centerline{\includegraphics[width=1\textwidth]{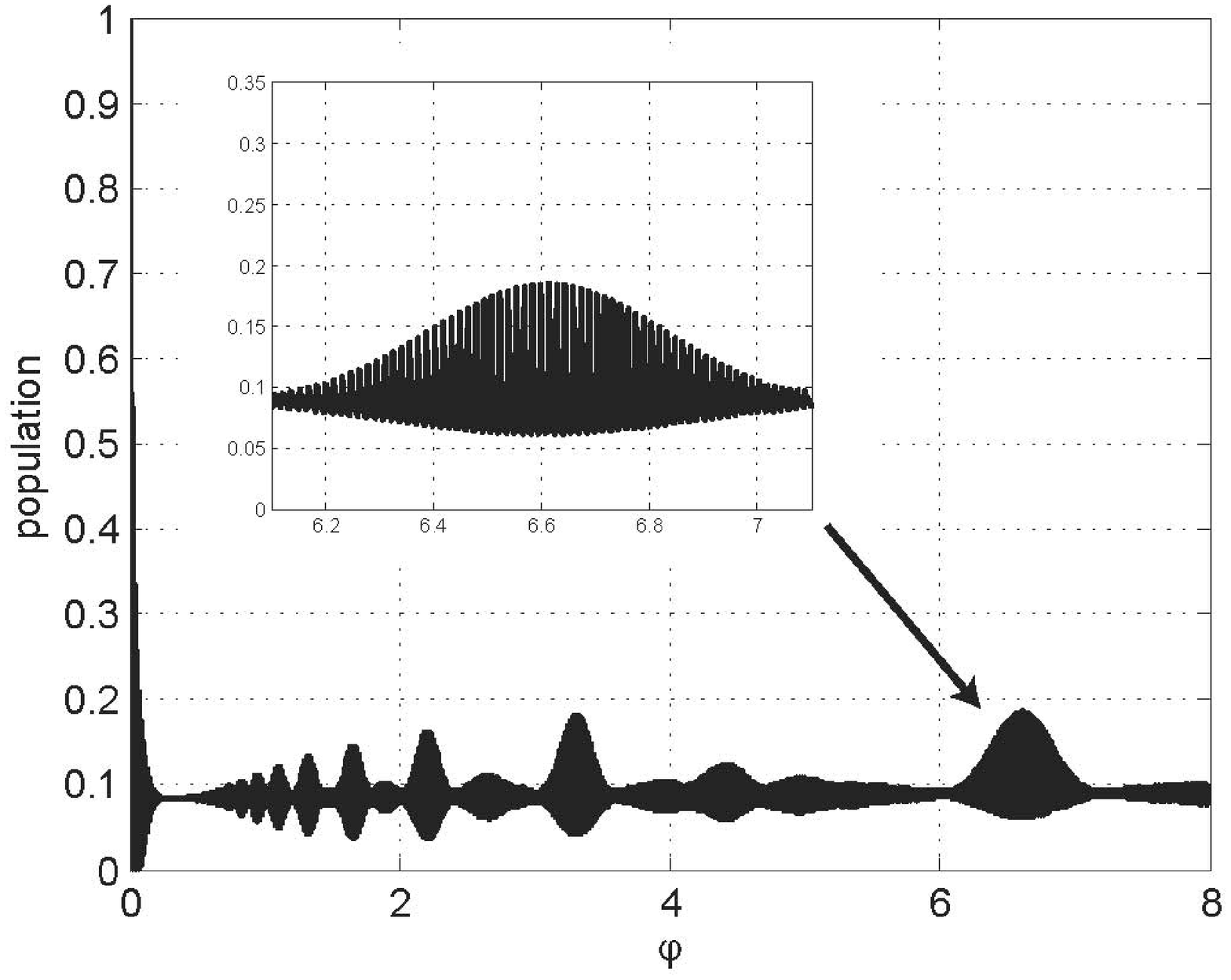}} % simu time 24
  \caption{Oscillation revival for $N_a=50$ and a coherent initial field
    with $\bar n=200$ photons; numerical solution of  rank $m=12$
    approximation~(\ref{eq:UH},\ref{eq:sigH}); we show the excited
    atomic  population (all atoms in the same  excited state) versus the
    adimensionalized time $\phi=\Omega_0 t/2\sqrt{\bar n}$; the
    parameter~$\kappa=\log(2) \Omega_0/(4\pi\bar n^{3/2})$ is adjusted
    according to~\cite{Meunier-et-al:PRA2006} in order to get a
    reduction  by a factor
    $2$ of the revival oscillations around $\phi=2\pi$. That reduction is indeed observed numerically.  }\label{fig:NspinRank12}
\end{figure}

\begin{figure}
\vspace*{-2cm}
\centerline{\includegraphics[width=.9\textwidth]{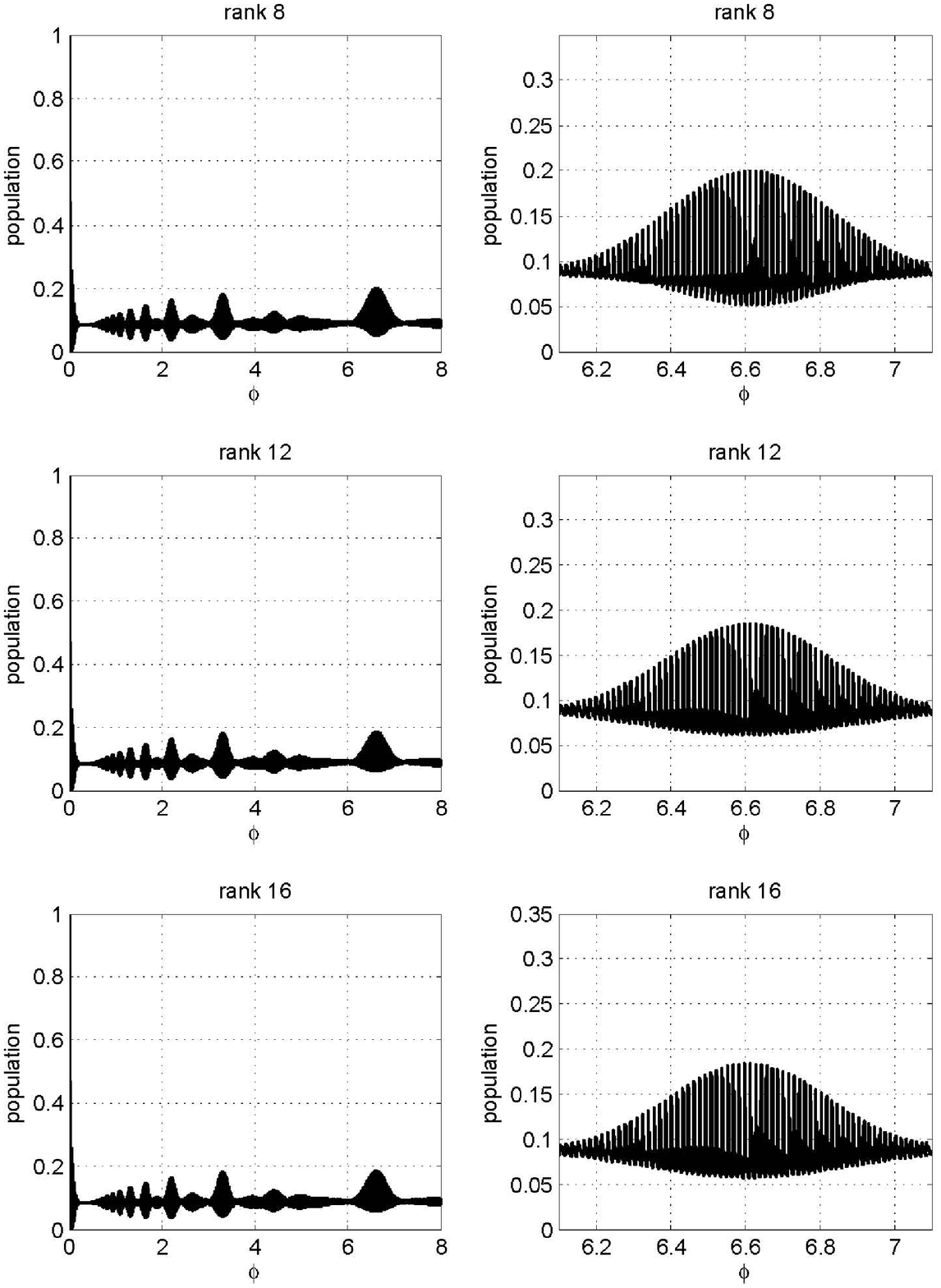}}
  \caption{Oscillation revival for $N_a=50$ and a coherent initial field
    with $\bar n=200$ photons; numerical solution of  rank $m=8,12,16$
    approximations~(\ref{eq:UH},\ref{eq:sigH}); we show the excited
    atomic  population (all atoms in the same  excited state) versus the
    adimensionalized time $\phi=\Omega_0 t/2\sqrt{\bar n}$; the
    parameter~$\kappa=\log(2) \Omega_0/(4\pi\bar n^{3/2})$ is adjusted
    according to~\cite{Meunier-et-al:PRA2006} in order to get a
    reduction  by a factor
    $2$ of the revival oscillations around $\phi=2\pi$.   }\label{fig:NspinPopRank8_16}
\end{figure}

\begin{figure}
\vspace*{-2cm}
\centerline{\includegraphics[width=.9\textwidth]{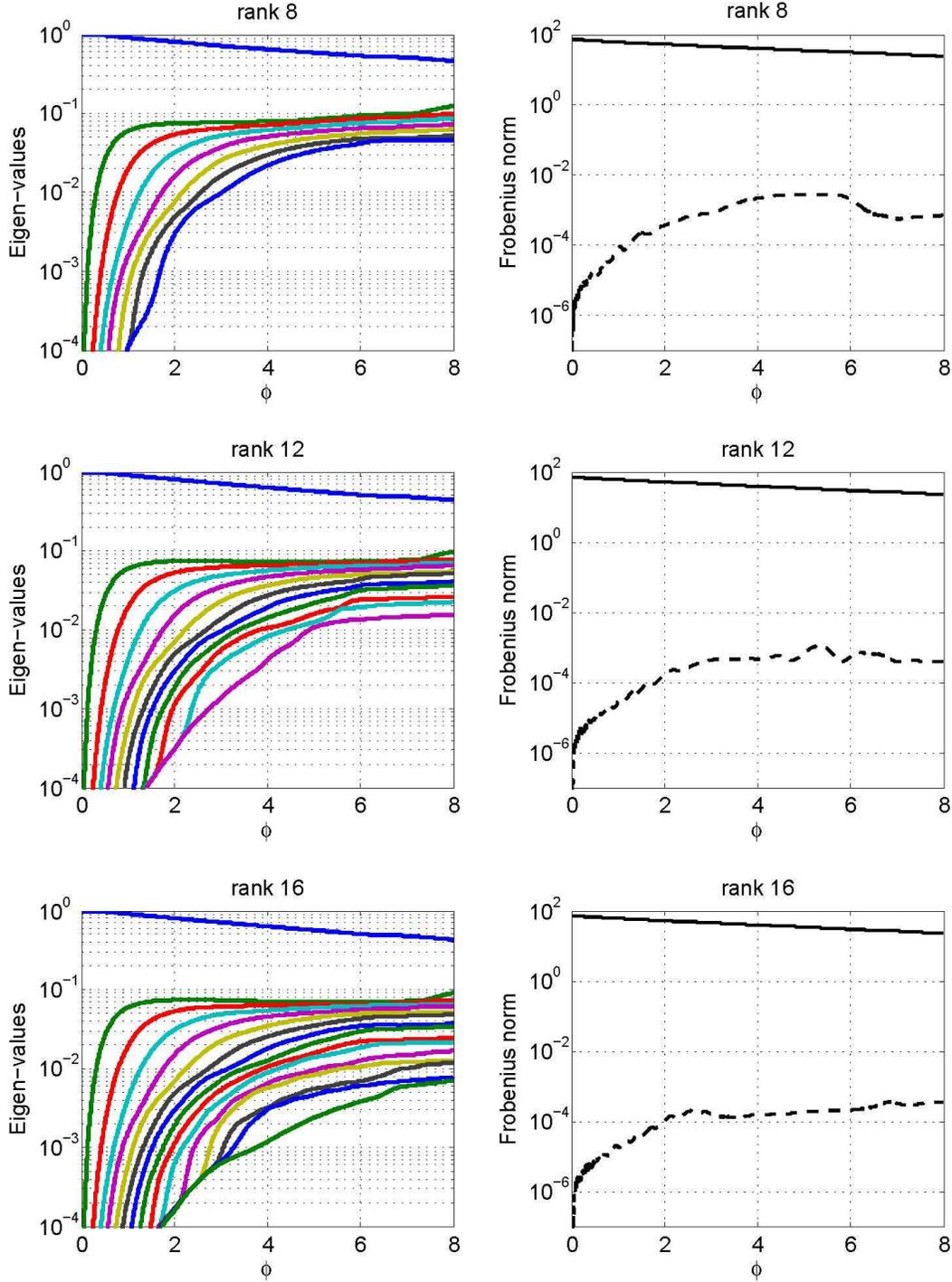}}
  \caption{Oscillation revival for $N_a=50$ and a coherent initial field
    with $\bar n=200$ photons; numerical solution of  rank $m=8,12,16$
    approximations~(\ref{eq:UH},\ref{eq:sigH}); left column eigenvalues of the approximate density matrices versus the
    adimensionalized time $\phi=\Omega_0 t/2\sqrt{\bar n}$;  right column Frobenius norms of $\dot\rho$ (solid line) and $\dot\rho_\perp$ (dashed line)  as explained in~\eqref{eq:rhoperp}; the other simulations parameters are identical to  figure~\ref{fig:NspinPopRank8_16}.
       }\label{fig:NspinSpectreRank8_16}
\end{figure}

 For $N_a=50$ atoms initially in the  same excited state and a  field
 initially in a coherent state with $\bar n=200$ photons a direct
 numerical integration of the Lindblad master equation~\eqref{eq:Jaynes-Cummings-Natoms}, when $\kappa >0$, is
 impossible on the limited computing facilities we have access to. The size $n$ of any
 finite dimensional approximation of $\rho$ by an  $n\times n$ matrix
 necessarily  exceeds $10000$.   For the purpose of validating our
 reduced model, we therefore proceed otherwise. With a  truncation to $300$ photons
 ($n=51\times 301$), we report here two  simulations.
On~Figure~\ref{fig:Nspin}, we show a reference simulation that is  a
  simulation of the exact Schr\"{o}dinger equation: the
  parameter~$\kappa$ in~\eqref{eq:Jaynes-Cummings-Natoms} is put to zero. The complete revival
  is maximum. On Figure~\ref{fig:NspinRank12}, we simulate our reduced
  model,  based on~(\ref{eq:UH},\ref{eq:sigH}) with  rank $m=12$, for the
  parameter value $\kappa=
 \log(2) \Omega_0/(4\pi\bar n^{3/2})$.  This specific choice of $\kappa >0$
 corresponds to a theoretical reduction by a factor 2 of the complete
 revival appearing at the adimensionalized time~$\phi=\Omega_0 t/2\sqrt{\bar n}=2\pi$   as
 predicted by the formula~(45) of~\cite{Meunier-et-al:PRA2006}.
 Our simulation is  performed with  the numerical scheme described in
 Section~\ref{sec:num} with a timestep $\dt=1/(\Omega_0\sqrt{\bar n}
 N_a)$.  Comparing
 Figure~\ref{fig:NspinRank12} to Figure~\ref{fig:Nspin}, we recover
 numerically that the revival amplitude around $\phi=2\pi$  is
 indeed reduced by a factor $2$. We have also compared in figures~\ref{fig:NspinPopRank8_16} and~\ref{fig:NspinSpectreRank8_16} rank $m=12$ with rank $m=8$ and rank $m=16$ approximations. On figure~\ref{fig:NspinPopRank8_16} we see that increasing the rank $m$ to 16 does not change the revival amplitude whereas decreasing the rank $m$ to 8 yields a slightly larger revival. On figure~\ref{fig:NspinSpectreRank8_16} we observe that the largest 8   eigenvalues in rank $m=12$ and rank $m=16$ simulations almost coincide with the 8 eigenvalues of rank $m=8$ simulations.  This clearly validates our reduced
 model. To complete this validation, we have checked that  smaller timestep and an higher rank $m=20$  do not significantly change  the numerical results. For
 completeness, we mention that the  large scale numerical results of Figures~\ref{fig:Nspin} and~\ref{fig:NspinRank12}  come from
 computations performed using a simple Matlab code available  upon
 request from the second author. The computations  were  executed on a Dell
 Precision M4400  computer equipped with Intel(R) Core(TM) duo CPU T9600
 at 2.80GHz with 8.00 Go RAM.  The rank $m=12$ simulation of
 Figure~\ref{fig:NspinRank12} typically takes about 24 hours.

\clearpage
 \section{Conclusion}
\label{sec:conclusion}

 For the numerical integration of  high dimensional open quantum
 systems, we have  proposed an approximation approach based on an
 orthogonal projection onto the set of  rank $m$ density matrices. A
 numerical integration scheme of the system of equations obtained for
 the reduced, low-rank model has been suggested. Its  derivation has been
 specifically adapted to the situation where the Hamiltonian evolution
 is must faster than the decoherence evolution. Other situations could
 be similarly investigated. The scheme has been implemented and tested
 on   oscillation revivals of atoms interacting resonantly  with a
 slightly damped  coherent quantized field. The results obtained show
 the good quality of the approximation, along with an evident
 significant speed-up with respect to the direct simulation of the full
 system. Although definite conclusions are yet to be obtained on cases
 of much larger sizes and extensive comparisons with the classically employed
 Monte-Carlo approaches are yet to be performed, the satisfactory
 results obtained to date show the promising nature of the
 approach. We additionally note that the strategy of approximation
 developed here is not restricted to open quantum systems. As shown in the appendix, it can be easily  adapted to deal with matrix Riccati equations appearing in  Kalman filtering, learning and  data fusion problems.

%\bibliographystyle{plain}
%\bibliography{E:/Latex/rhn}

\appendix
\section{The Low-rank  differential  Riccati  equation} \label{ap:riccati}
This appendix relies on an  adaptation  of the lifting method used in~\cite{bonnabel-sepulchre:2012}. We compute here  the lift of the orthogonal projection onto the rank $m$ covariance matrices of the vector-field defined by the matrix Riccati equation.

 We consider  the simplest situation: the state $x\in\RR^n$ and output $y\in\RR^p$ are related by the stochastic differential equations ($dw\in\RR^n$ and $d\eta\in \RR^d$ are vectors  whose components are independent Wiener processes of standard deviation $1$)
 $$ d x = Ax~dt + G~dw \quad\text{and}\quad  dy = C~dx + H~d\eta
 $$
 where $A$, $G$, $C$ and $H$ are respectively $n\times n$, $n\times n$, $p\times n$ and  $p\times p$  matrices with real entries. When $H$ is invertible,  the
 computation of the best estimate of $x$ at $t$ knowing the past values of the output $y$ relies on the computation of the conditional state-error covariance matrix $P$ solution to  the Riccati matrix  equation
\begin{equation}\label{eq:riccati}
\dotex P = A P + P A^T + G G^T - P C^T (HH^T)^{-1} C P
\end{equation}
where $^T$ stands for transpose. The symmetric $n\times n$ matrix $P$ is non-negative.

When $G=0$, this Riccati equation is rank preserving. It defines then  a vector field on the sub-manifold of rank $m < n$ covariance matrices among the symmetric  $n\times n$ matrices with real entries and equipped with the Frobenius scalar product. This sub-manifold denoted by $\mathcal{P}_m$  admits the over-parameterization
$$
(O,R) \mapsto O R O^T=P
$$
where $O$ belongs to the set of $n\times m$ orthogonal matrices ($O^T O = \Id_m$) and $R$ is $m\times m$,  positive definite and   symmetric.
We will prove here that the analogue of~\eqref{eq:U} and~\eqref{eq:sig}  reads:
\begin{align}
    \label{eq:O}
    \dotex O &= \Omega O +(\Id_n-O O^T)\big( (A-\Omega) O + G G^T O R^{-1} \big)
    \\
    \dotex R &= O^T (A-\Omega) O R + R O^T(A^T+\Omega) O + O^T GG^T O - R O^TC^T (HH^T)^{-1} C O R
     \label{eq:R}
\end{align}
where $\Omega$ is an arbitrary skew-symmetric $n\times n$ matrix ($\Omega^T=-\Omega$) playing the role of gauge degrees of freedom. The matrix $\Omega$ could possibly depend on $t$, $O$ or  $R$. With $\Omega=0$ and $G=0$ we recover the lift given in~\cite{bonnabel-sepulchre:2012}:
$$
\dotex O = (\Id_n - O O^T) A O, \quad \dotex R= O^T A O R + R O^T A O - R O^T C (HH^T)^{-1} C O R
.
$$
Notice that we can adapt the discretization strategy used for~\eqref{eq:U} and~\eqref{eq:sig} to get with a proper choice of the gauge matrix $\Omega$ an efficient numerical scheme. Let us conclude by proving that~\eqref{eq:O} and~\eqref{eq:R} correspond to a lift of the orthogonal projection onto  $\mathcal{P}_m$  of the vector-field defined by~\eqref{eq:riccati}.

Take $P=ORO^T$ a rank $m$ covariance matrix. The tangent space at $P$ to $\mathcal{P}_m$ is parameterized by $\dO R O^T +O \dR O^T + O R \dO^T$ where $\dO=\omega O $ and $\dR=\xi$ are arbitrary variations of $O$ and $R$ associated to any $\omega$ and $\xi$,  skew-symmetric $n\times n$  and symmetric $m\times m$ matrices. The orthogonal projection of $\dotex P$ given by~\eqref{eq:riccati} onto this tangent space is associated to $\omega$ and $\xi$ minimizing the Frobenius distance between $\dO R O^T +O \dR O^T + O R \dO^T $ and $A P + P A^T + G G^T - P C^T (HH^T)^{-1} C P$. This means that the skew-symmetric matrix $\omega$ and the symmetric matrix  $\xi$ minimize
$$
\tr{\left(A P + P A^T + G G^T - P C^T (HH^T)^{-1} C P - (\omega P - P\omega + O\xi O^T) \right)^2}
.
$$
The first order stationary conditions characterizing $\omega$ and $\xi$ read
\begin{align*}
   0&= O^T \big(A P + P A^T + G G^T - P C^T (HH^T)^{-1} C P - (\omega P - P\omega + O\xi O^T) \big) O
   \\
   0&=\Big[ P\quad , \quad A P + P A^T + G G^T - P C^T (HH^T)^{-1} C P - (\omega P - P\omega + O\xi O^T) \Big]
   .
\end{align*}
Since $O^TO=\Id_m$ we have
\begin{equation}\label{eq:xi}
\xi =O^T \big(A P + P A^T + G G^T - P C^T (HH^T)^{-1} C P -\omega P + P\omega) \big) O
\end{equation}
and thus $\omega$ is solution to the following matrix equation
$$
    (\Id_n-OO^T) \big( A P +G G^T - \omega P \big)P
=
P \big(P A^T   + G G^T + P\omega \big) (\Id_n-OO^T)
$$
obtained after some manipulations using  $O^T O = \Id_m$ and $P(\Id_n-OO^T)=(\Id_n-OO^T) P = 0$
This matrix equation admits the following general solution
\begin{multline}\label{eq:omega}
   \omega = (\Id_n-OO^T) \big( AO + GG^TOR^{-1}\big) O^T -O \big( O^TA^T + R^{-1} O^T GG^T\big)  (\Id_n-OO^T)
    \\
    + OO^T\Omega OO^T + (\Id_n-OO^T)\Omega(\Id_n-OO^T)
\end{multline}
where $\Omega$ is an  arbitrary skew-symmetric matrix.
We get~\eqref{eq:O} since
$$
\omega O = \Omega O +(\Id_n-O O^T)\big( (A-\Omega) O + G G^T O R^{-1} \big)
.
$$
To get~\eqref{eq:R} we  take  $\omega$ given by~\eqref{eq:omega}  and insert  it into~\eqref{eq:xi}.

\end{document}